# Multi-party quantum private comparison of size relationship with two third parties based on $d$-dimensional Bell states

Jiang-Yuan Lian, Xia Li, Tian-Yu Ye*

College of Information & Electronic Engineering, Zhejiang Gongshang University, Hangzhou 310018, P.R. China

E-mail: yetianyu@zjgsu.edu.cn (T.Y. Ye)

**Abstract:** In this paper, we put forward a multi-party quantum private comparison (MQPC) protocol with two semi-honest third parties (TPs) by adopting $d$-dimensional Bell states, which can judge the size relationship of private integers from more than two users within one execution of protocol. Each TP is permitted to misbehave on her own but cannot collude with others. In the proposed MQPC protocol, TPs are only required to apply $d$-dimensional single-particle measurements rather than $d$-dimensional Bell state measurements. There are no quantum entanglement swapping and unitary operations required in the proposed MQPC protocol. The security analysis validates that the proposed MQPC protocol can resist both the outside attacks and the participant attacks. The proposed MQPC protocol is adaptive for the case that $n$ users want to compare the size relationship of their private integers under the control of two supervisors. Furthermore, the proposed MQPC protocol can be used in the strange user environment, because there are not any communication and pre-shared key between each pair of users.

## 1 Introduction

Classical private comparison is one of the most important branches of classical secure multiparty computation (SMC). In 1982, the first classical private comparison protocol, called as the millionaire problem, was proposed by Yao [1], which aims to determine who is the richer one within the mentioned two millionaires without leaking their actual wealth. However, the security of classical private comparison relies on the computation complexity, which means that the corresponding protocol cannot guarantee the security as long as the computer has enough ability to deal with extremely complex data. In order to overcome this problem, quantum private comparison (QPC) was put forward in 2009 by combining quantum mechanics and classical private comparison [2]. Subsequently, lots of two-party QPC protocols of equality, which can compare the equality of private inputs from two users, have been proposed [3-19]. In 2013, the first multi-party quantum private comparison (MQPC) protocol of equality, which can compare the equality of private inputs from more than two users within one execution of protocol, was put forward by Chang *et al.* [20]. Later, numerous MQPC protocols of equality have been constructed with different quantum technologies [21-26]. However, none of the QPC protocols in Refs.[2-26] have the function of judging the size relationship (i.e., greater than, equal to or smaller than) of private inputs from more than two users within one execution of protocol, which restricts their applications in practice.

In 2013, the first QPC protocol of size relationship was put forward by Lin *et al.* [27] by using $d$-dimensional Bell states, which can compare the size relationship of private inputs from two users. Then, Yu *et al.* [28] designed a two-party QPC protocol of size relationship with $d$-dimensional single-particle states in 2013; Guo *et al.* [29] proposed a two-party QPC protocol of size relationship based on entanglement swapping of $d$-dimensional Bell states in 2014; Chen *et al.* [30] put forward a two-party QPC protocol of size relationship via quantum walks on circle in 2021. However, the QPC protocols of size relationship in Refs.[27-30] are not feasible for the multi-user circumstance. Fortunately, in 2014, Luo *et al.* [31] put forward the first MQPC protocol of size relationship by using $d$-dimensional entangled states, which can judge the size relationship of private inputs from more than two users within one execution of protocol. Since then, a series of MQPC protocols of size relationship have been proposed, such as the ones with single-particle states [32-33], the one with $d$-dimensional GHZ states [34] and the one with $d$-dimensional Bell states [35]. At present, the number of MQPC protocols of size relationship is still few. Moreover, the first protocol of Ref.[32] is the only MQPC protocol of size relationship which has two supervisors.

Based on the above analysis, in this paper, we are devoted to considering the case that $n$ strange users want to compare the size relationship of their private integers under the control of two supervisors. We use $d$-dimensional Bell states to design a novel MQPC protocol of size relationship with two semi-honest third parties (TPs). Two TPs only need to perform $d$-dimensional single-particle measurements rather than $d$-dimensional Bell state measurements. The proposed MQPC protocol requires neither quantum entanglement swapping nor unitary operations. The proposed MQPC protocol can be adaptive for the strange user environment.

## 2 Protocol description

A $d$-dimensional Bell state can be depicted as

$$|\phi_{u,v}\rangle = \frac{1}{\sqrt{d}} \sum_{j=0}^{d-1} e^{\frac{2\pi i j u}{d}} |j\rangle |j \oplus v\rangle, \tag{1}$$

where $u, v \in \{0, 1, \ldots, d-1\}$, and $\oplus$ denotes the modulo $d$ addition. Two common conjugate bases in the $d$-dimensional quantum system can be described as

$$T_1 = \{|0\rangle, |1\rangle, \ldots, |d-1\rangle\}, \tag{2}$$

$$T_2 = \{F|0\rangle, F|1\rangle, \ldots, F|d-1\rangle\}, \tag{3}$$

where $F|t\rangle = \frac{1}{\sqrt{d}} \sum_{\alpha=0}^{d-1} e^{\frac{2\pi i \alpha t}{d}} |\alpha\rangle$ with $t = 0, 1, \ldots, d-1$ represents the $d$-dimensional discrete quantum Fourier transform. It is apparent that, when the two qudits of the $d$-dimensional Bell state in Eq.(1)

are measured with the $T_1$ basis, they are collapsed into $|j\rangle$ and $|j\oplus v\rangle$, respectively. As a result, we have $j\oplus v\ominus j=v$, where $\ominus$ denotes the modulo $d$ subtraction. We will use this property of the $d$-dimensional Bell state in Eq.(1) to design the proposed MQPC protocol.

Suppose that there are $n$ strange users, $P_1, P_1, \ldots, P_n$, whose private integers can be represented by $p_1, p_2, \ldots, p_n$, respectively; $TP_1$ and $TP_2$ are two TPs, each of whom is permitted to misbehave on her own but cannot collude with others. Here, $p_i \in \{0,1,\ldots,h\}$ and $i=1,2,\ldots,n$. When $mod(d,2)=0$, we set $h=\frac{d}{2}$; otherwise, we set $h=\frac{d-1}{2}$. $P_i$ and $TP_2$ pre-share the private key $k_i$ through a secure quantum key distribution (QKD) protocol, $k_i \in \{0,1,\ldots,d-1\}$ and $i=1,2,\ldots,n$. $P_1, P_1, \ldots, P_n$ want to compare the size relationship of their private integers under the control of $TP_1$ and $TP_2$ on the condition that there are not any communication and pre-shared key between any two of them, which implies that only when both $TP_1$ and $TP_2$ agree can $P_1, P_1, \ldots, P_n$ obtain the size relationship of their private integers. In the following, we put forward a novel MQPC protocol suitable for the strange user environment to accomplish this goal. Here, the quantum channels are assumed to be noiseless, while the classical channels are supposed to be authenticated.

Step 1: $TP_1$ prepares $n$ $d$-dimensional Bell states in Eq.(1), and picks out all of the first and the second particles of these $n$ Bell states to form sequences $S_1$ and $S_2$, respectively. Here, $S_1$ and $S_2$ can be represented by

$$S_1 = \left[ |m_1^1\rangle, |m_1^2\rangle, \ldots, |m_1^n\rangle \right] \tag{4}$$

and

$$S_2 = \left[ |m_2^1\rangle, |m_2^2\rangle, \ldots, |m_2^n\rangle \right], \tag{5}$$

respectively. Then, $TP_1$ records the second label of the $i$ th $d$-dimensional Bell state as $v_i$. Here, $m_1^i, m_2^i \in \{0,1,\ldots,d-1\}$ and $i=1,2,\ldots,n$.

Step 2: $TP_1$ prepares $L$ decoy photons which are randomly selected from the sets $T_1$ and $T_2$. Then she inserts them into $S_1$ to form a new sequence $S_1'$ and sends it to $TP_2$ via a quantum channel. After all particles of $S_1'$ are received by $TP_2$, $TP_1$ and $TP_2$ start the security check on the quantum channel between them. To be specific, $TP_1$ announces the positions and the preparation bases of all decoy photons in $S_1'$ to $TP_2$; then, $TP_2$ acquires the measurement results of these decoy photons by measuring them with the correct measuring bases and returns these measurement results to $TP_1$; afterward, $TP_1$ judges whether there is an eavesdropper or not by comparing the received measurement results with decoy photons' initial prepared states. If there is an eavesdropper, the protocol will be aborted; otherwise, the protocol will be proceeded.

Step 3: $TP_1$ prepares $n$ groups of $L$ decoy photons which are randomly selected from the sets $T_1$ and $T_2$. Then, $TP_1$ inserts the particle $|m_2^i\rangle$ into the $i$ th decoy photon group at random position to construct a new sequence $G_i$, where $i = 1, 2, \ldots, n$. Finally, $TP_1$ sends $G_i$ to $P_i$.

Step 4: $TP_1$ announces the positions and the preparation bases of all decoy photons to $P_i$, where $i = 1, 2, \ldots, n$. Based on published positions, $P_i$ measures the decoy photons with the correct measuring bases and publishes the corresponding measurement results to $TP_1$. Afterward, $TP_1$ judges whether the quantum channel is secure or not by comparing the received measurement results with the original states of decoy photons. If the quantum channel is secure, the protocol will be proceeded; otherwise, the protocol will be terminated.

Step 5: $P_i$ measures the particle $|m_2^i\rangle$ with the $T_1$ basis and records the value $m_2^i$, where $i = 1, 2, \ldots, n$. Then, $P_i$ calculates

$$r_2^i = m_2^i \oplus p_i \oplus k_i. \quad (6)$$

After that, $P_i$ sends $r_2^i$ to $TP_2$ via an authenticated classical channel.

Step 6: $TP_2$ measures the $n$ particles $|m_1^1\rangle, |m_1^2\rangle, \ldots, |m_1^n\rangle$ with the $T_1$ basis and records the values $m_1^1, m_1^2, \ldots, m_1^n$. Then, $TP_2$ generates a private integer $q$, where $q \in \{h, h+1, \ldots, d-1\}$. Afterward, $TP_2$ computes

$$r_1^i = m_1^i \oplus q \oplus k_i, \quad (7)$$

where $i = 1, 2, \ldots, n$. After that, $TP_2$ makes

$$r_i = r_1^i \ominus r_2^i. \quad (8)$$

Finally, $TP_2$ sends $R$ to $TP_1$ via an authenticated classical channel, where $R = [r_1, r_2, \ldots, r_n]$.

Step 7: $TP_1$ calculates

$$M_i = (d-1) - (r_i \oplus v_i), \quad (9)$$

where $i = 1, 2, \ldots, n$. Here, the size relationship of different $M_i$ s is same to that of different $p_i$ s. Finally, $TP_1$ announces the size relationship of different $p_i$ s to $P_1, P_1, \ldots, P_n$ through an authenticated classical channel.

In addition, in order to make the process of the proposed MQPC protocol easier to understand, we draw it in Fig.1 after ignoring the procedures of security check.

# 3 Correctness analysis

## 3.1 Formula analysis

Based on Eq.(1) and entanglement property of two qudits within a Bell state, we can infer

$$m_1^i \ominus m_2^i \oplus v_i = 0 . \tag{10}$$

According to Eqs.(6-8) and Eq.(10), we can calculate

$$\begin{aligned} r_i \oplus v_i &= r_1^i \ominus r_2^i \oplus v_i \\ &= (m_1^i \oplus q \oplus k_i) \ominus (m_2^i \oplus p_i \oplus k_i) \oplus v_i \\ &= (m_1^i \oplus q) \ominus (m_2^i \oplus p_i) \oplus v_i \\ &= \left( m_1^i \ominus m_2^i \oplus v_i \right) \oplus q \ominus p_i \\ &= q \ominus p_i . \end{aligned} \tag{11}$$

By virtue of Eq.(9) and Eq.(11), we can obtain

$$M_i = (d-1) - (q \ominus p_i) . \tag{12}$$

Since $p_i \in \{0,1,\ldots,h\}$ and $q \in \{h, h+1, \ldots, d-1\}$, we have $q \ominus p_i = q - p_i \in \{0,1,\ldots,d-1\}$. Therefore, we get

$$M_i = p_i + d - 1 - q . \tag{13}$$

Consequently, it can be concluded that the size relationship of different $M_i$ s is same to that of different $p_i$ s.

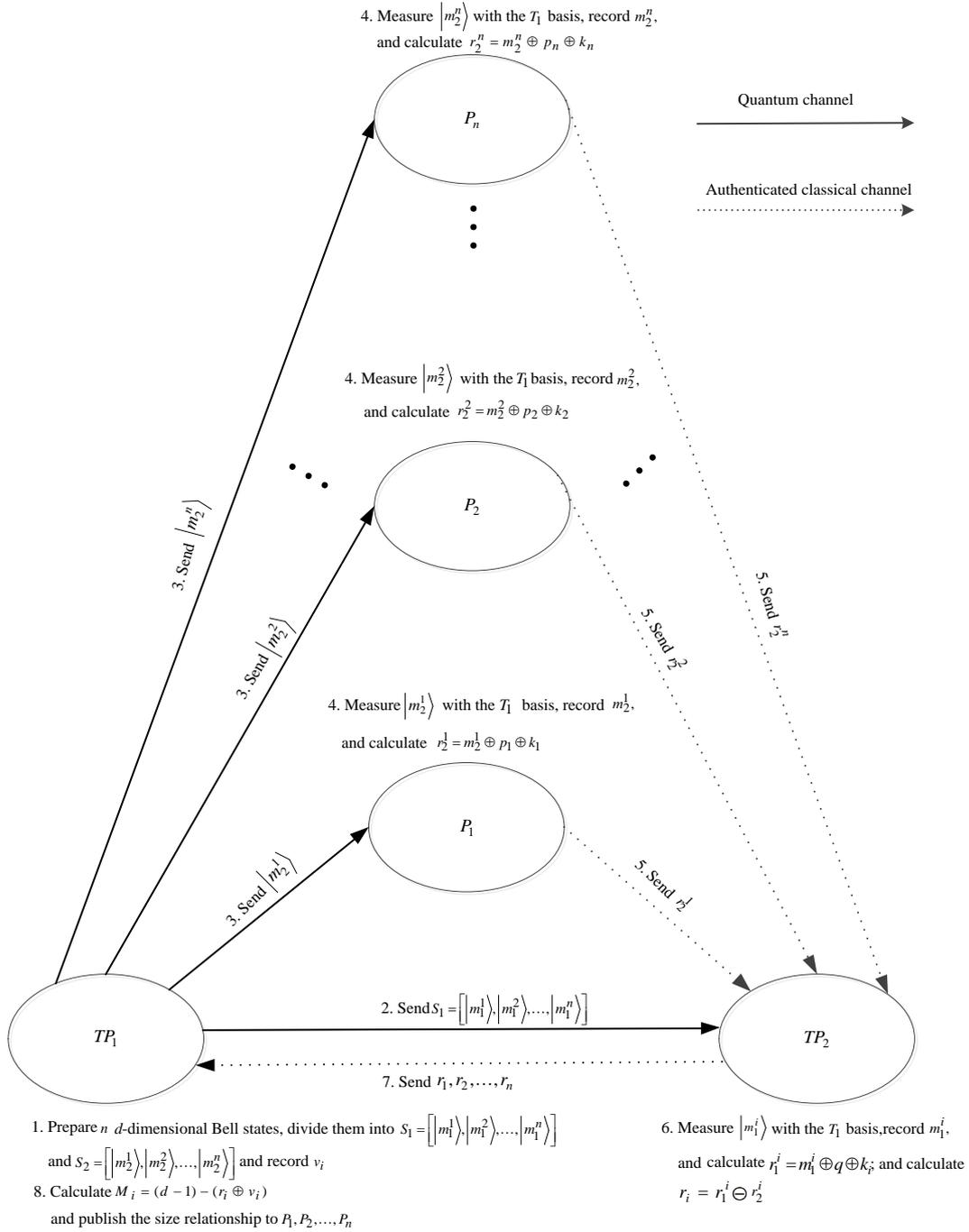

Fig.1 The flow chart of the proposed MQPC protocol

### 3.2 Examples

In order to further prove the correctness of the proposed MQPC protocol, we show a specific example in detail. Suppose that there are four users, $P_1, P_2, P_3, P_4$, whose private integers are $p_1 = 4, p_2 = 3, p_3 = 1, p_4 = 5$, respectively; $TP_1$ prepares four 11-dimensional Bell states and makes $v_1 = 3, v_2 = 4, v_3 = 5, v_4 = 6$; $P_1, P_2, P_3, P_4$ pre-share the private keys $k_1 = 7, k_2 = 8, k_3 = 6, k_4 = 2$ with $TP_2$, respectively; $TP_2$ measures the four particles $|m_1^1\rangle, |m_1^2\rangle, |m_1^3\rangle, |m_1^4\rangle$ with the $T_1$ basis and obtains $m_1^1 = 6, m_1^2 = 9, m_1^3 = 8, m_1^4 = 3$, respectively; $P_1, P_2, P_3, P_4$ measure the four particles

$\left|m_2^1\right\rangle,\left|m_2^2\right\rangle,\left|m_2^3\right\rangle,\left|m_2^4\right\rangle$ with the $T_1$ basis and obtain $m_2^1=9, m_2^2=2, m_2^3=2, m_2^4=9$, respectively; and $TP_2$ generates a private integer $q=6$.

According to Eq.(6), $P_1, P_2, P_3, P_4$ can get $r_2^1 = 9 \oplus 4 \oplus 7 = 9$, $r_2^2 = 2 \oplus 3 \oplus 8 = 2$, $r_2^3 = 2 \oplus 1 \oplus 6 = 9$ and $r_2^4 = 9 \oplus 5 \oplus 2 = 5$, respectively. According to Eq.(7), $TP_2$ can obtain $r_1^1 = 6 \oplus 6 \oplus 7 = 8$, $r_1^2 = 9 \oplus 6 \oplus 8 = 1$, $r_1^3 = 8 \oplus 6 \oplus 6 = 9$ and $r_1^4 = 3 \oplus 6 \oplus 2 = 0$. By virtue of Eq.(8), $TP_2$ can acquire $r_1 = 8 \ominus 9 = 10$, $r_2 = 1 \ominus 2 = 10$, $r_3 = 9 \ominus 9 = 0$ and $r_4 = 0 \ominus 5 = 6$. Based on Eq.(9), $TP_1$ can obtain $M_1 = 10 - (10 \oplus 3) = 8$, $M_2 = 10 - (10 \oplus 4) = 7$, $M_3 = 10 - (0 \oplus 5) = 5$ and $M_4 = 10 - (6 \oplus 6) = 9$. It is obvious that $M_4 > M_1 > M_2 > M_3$, so we can infer that $p_4 > p_1 > p_2 > p_3$.

## 4 Security analysis

### 4.1 External attacks

(1) The intercept-resend attack

In order to acquire some useful information, an outside eavesdropper, Eve, may intercept the particles of $S_1^{'}$ ($G_i$) sent out from $TP_1$ in Step 2 (3) and replace them with the fake ones which are prepared beforehand by herself randomly in the $T_1$ basis or the $T_2$ basis. Unfortunately, Eve's fake particles are not necessarily the same as the genuine ones, which makes her attack behavior undoubtedly discovered. Concretely speaking, for one decoy photon, the probability that Eve's attack on it can be detected is $\frac{d-1}{d}$; for $L$ decoy photons, the probability that Eve's attacks on them can be detected is $1-\left(1-\frac{d-1}{d}\right)^L = 1-\left(\frac{1}{d}\right)^L$, which converges to 1 when $L$ is large enough.

(2) The measure-resend attack

In order to acquire some useful information, Eve may intercept the particles of $S_1^{'}$ ($G_i$) sent out from $TP_1$ in Step 2 (3), measure them randomly with the $T_1$ basis or the $T_2$ basis and sends the resulted states to $TP_2$ ($P_i$). Because Eve's measuring basis on decoy photons are not always consistent with their prepared basis, she inevitably leaves her trace on decoy photons. Concretely speaking, for one decoy photon, Eve's attack on it can be discovered with the probability of $\frac{d-1}{2d}$ in Step 2 (3); hence, for $L$ decoy photons, Eve's attacks on them can be detected with the probability of $1-\left(1-\frac{d-1}{2d}\right)^L = 1-\left(\frac{d+1}{2d}\right)^L$, which approaches 1 when $L$ is large enough.

(3) The entangle-measure attack

Eve may perform the entangle-measure attack on the particles of $S_1^{'}$ sent from $TP_1$ to $TP_2$ in Step 2.

**Theorem.** *Suppose that Eve imposes her unitary operation $U_E$ on the particle of $S_1^{'}$ from $TP_1$ to $TP_2$ and her probe particle $|\zeta\rangle$. For inducing no error in Step 2, the final state of Eve's probe should* be independent from the particle of $S_1^{'}$ from $TP_1$ to $TP_2$. Consequently, Eve gets nothing useful.

**Proof.** Firstly, consider the case that the particle of $S_1^{'}$ is produced by $TP_1$ in the $T_1$ basis. We use $|t\rangle$ to represent the particle of $S_1^{'}$ in the $T_1$ basis. After Eve performs $U_E$ on $|t\rangle$ and $|\varsigma\rangle$, the global composite quantum system is turned into

$$U_E(|t\rangle|\varsigma\rangle) = \sum_{t'=0}^{d-1} \beta_{tt'} |t'\rangle |\varsigma_{tt'}\rangle, \qquad (14)$$

where $|\varsigma_{tt'}\rangle$ ($t, t' = 0, 1, \ldots, d-1$) are Eve's probe states dependent on $U_E$, and

$$\sum_{t'=0}^{d-1} |\beta_{tt'}|^2 = 1. \qquad (15)$$

In order for Eve to escape the eavesdropping detection in Step 2, it should satisfy that $\beta_{tt'} = 0$ for $t \neq t'$. As a result, Eq.(14) is changed into

$$U_E(|t\rangle|\varsigma\rangle) = \beta_{tt} |t\rangle |\varsigma_{tt}\rangle \qquad (16)$$

for $t = 0, 1, \ldots, d-1$.

Secondly, consider the case that the particle of $S_1^{'}$ is generated by $TP_1$ in the $T_2$ basis. For simplicity, we use $|J_t\rangle = F|t\rangle = \frac{1}{\sqrt{d}} \sum_{\alpha=0}^{d-1} e^{\frac{2\pi i \alpha t}{d}} |\alpha\rangle$ to represent the particle of $S_1^{'}$ in the $T_2$ basis, where $t = 0, 1, \ldots, d-1$. After Eve performs $U_E$ on $|J_t\rangle$ and $|\varsigma\rangle$, the global composite quantum system is transformed into

$$U_E(|J_t\rangle|\varsigma\rangle) = U_E \left[ \left( \frac{1}{\sqrt{d}} \sum_{\alpha=0}^{d-1} e^{\frac{2\pi i \alpha t}{d}} |\alpha\rangle \right) |\varsigma\rangle \right]$$

$$= \frac{1}{\sqrt{d}} \sum_{\alpha=0}^{d-1} e^{\frac{2\pi i \alpha t}{d}} U_E(|\alpha\rangle|\varsigma\rangle). \qquad (17)$$

After inserting Eq.(16) into Eq.(17), we have

$$U_E(|J_t\rangle|\varsigma\rangle) = \frac{1}{\sqrt{d}} \sum_{\alpha=0}^{d-1} e^{\frac{2\pi i \alpha t}{d}} \beta_{\alpha\alpha} |\alpha\rangle |\varsigma_{\alpha\alpha}\rangle. \qquad (18)$$

According to the inverse quantum Fourier transform, we can acquire

$$|\alpha\rangle = \frac{1}{\sqrt{d}} \sum_{\gamma=0}^{d-1} e^{-\frac{2\pi i \gamma \alpha}{d}} |J_\gamma\rangle. \qquad (19)$$

Inserting Eq.(19) into Eq.(18) produces

$$U_E(|J_t\rangle|\varsigma\rangle) = \frac{1}{\sqrt{d}} \sum_{\alpha=0}^{d-1} e^{\frac{2\pi i \alpha t}{d}} \beta_{\alpha\alpha} \left( \frac{1}{\sqrt{d}} \sum_{\gamma=0}^{d-1} e^{-\frac{2\pi i \gamma \alpha}{d}} |J_\gamma\rangle \right) |\varsigma_{\alpha\alpha}\rangle$$

$$= \frac{1}{d}\sum_{\alpha=0}^{d-1}\sum_{\gamma=0}^{d-1} e^{\frac{2\pi i\alpha(t-\gamma)}{d}}\beta_{\alpha\alpha}|J_\gamma\rangle|\zeta_{\alpha\alpha}\rangle$$

$$= \frac{1}{d}\left(|J_0\rangle\sum_{\alpha=0}^{d-1} e^{\frac{2\pi i\alpha(t-0)}{d}}\beta_{\alpha\alpha}|\zeta_{\alpha\alpha}\rangle + |J_1\rangle\sum_{\alpha=0}^{d-1} e^{\frac{2\pi i\alpha(t-1)}{d}}\beta_{\alpha\alpha}|\zeta_{\alpha\alpha}\rangle + \ldots + |J_{d-1}\rangle\sum_{\alpha=0}^{d-1} e^{\frac{2\pi i\alpha[t-(d-1)]}{d}}\beta_{\alpha\alpha}|\zeta_{\alpha\alpha}\rangle\right). \quad (20)$$

In order that Eve cannot be detected in Step 2, it should meet

$$\sum_{\alpha=0}^{d-1} e^{\frac{2\pi i\alpha(t-\gamma)}{d}}\beta_{\alpha\alpha}|\zeta_{\alpha\alpha}\rangle = 0 \quad (21)$$

for $t \neq \gamma$ and $t, \gamma = 0,1,\ldots,d-1$. When $t \neq \gamma$, we can get

$$\sum_{\alpha=0}^{d-1} e^{\frac{2\pi i\alpha(t-\gamma)}{d}} = 0. \quad (22)$$

Based on Eq.(21) and Eq.(22), we can acquire

$$\beta_{00}|\zeta_{00}\rangle = \beta_{11}|\zeta_{11}\rangle = \ldots = \beta_{(d-1)(d-1)}|\zeta_{(d-1)(d-1)}\rangle = \beta'|\zeta'\rangle. \quad (23)$$

Thirdly, by combing Eq.(16) and Eq.(23), we can get

$$U_E(|t\rangle|\zeta\rangle) = \beta'|t\rangle|\zeta'\rangle. \quad (24)$$

Inserting Eq.(21) and Eq.(23) into Eq.(20) generates

$$U_E(|J_t\rangle|\zeta\rangle) = \beta'|J_t\rangle|\zeta'\rangle. \quad (25)$$

According to Eq.(24) and Eq.(25), in order for introducing no error in Step 2, the final state of Eve's probe should be independent from the particle of $S_1'$ from $TP_1$ to $TP_2$. Consequently, Eve gets nothing useful by launching this kind of attack.

Besides, Eve may perform the entangle-measure attack on the particles of $G_i$ sent from $TP_1$ to $P_i$ in Step 3. We can also prove in a same way as above that Eve still cannot acquire any useful information by doing this.

### 4.2 Participant attacks

(1) The participant attack from one dishonest user

In the proposed MQPC protocol, $n$ users essentially play the equal roles. Without loss of generality, suppose that $P_1$ is the only dishonest user who tries to steal the private integers from other users.

In order to obtain something useful about $p_b$, where $b = 2,3,\ldots,n$, $P_1$ may launch the outside attacks on the particles of $S_1'$ sent out from $TP_1$ to $TP_2$ in Step 2 or the particles of $G_b$ sent out from $TP_1$ to $P_b$ in Step 3. Actually, $P_1$ essentially plays the same role as Eve. As a result, just as previously analyzed in Sect.4.1, $P_1$ cannot obtain something useful about $p_b$ without being detected.

Besides, $P_1$ may hear of $r_2^b$ from $P_b$ in Step 5. According to Eq.(6), $P_1$ cannot get $p_b$ since she has no way to acquire $m_2^b$ and $k_b$. In addition, $P_1$ may hear of $r_b$ from $TP_2$ in Step 6. According to Eq.(7) and Eq.(8), it is apparent that $P_1$ still has no way to obtain $p_b$. Finally, $P_1$ receives the size relationship of different $p_i$ s from $TP_1$ in Step 7, where $i = 1, 2, \ldots, n$. However, she still has no chance to acquire $p_b$.

In conclusion, one dishonest user has no chance to acquire other users' private integers.

(2) The participant attack from two or more dishonest users

We analyze the worst situation that the number of dishonest users is $n-1$. Without loss of generality, assume that the dishonest users are $P_1, P_2, \ldots, P_{a-1}, P_{a+1}, \ldots, P_n$, who try to steal the private integer $p_a$, where $a = 2, 3, \ldots, n-1$. $P_1, P_2, \ldots, P_{a-1}, P_{a+1}, \ldots, P_n$ may launch their attacks on the particles of $S_1'$ sent out from $TP_1$ to $TP_2$ in Step 2 or the particles of $G_a$ sent out from $TP_1$ to $P_a$ in Step 3. As described in Sect.4.1, $P_1, P_2, \ldots, P_{a-1}, P_{a+1}, \ldots, P_n$ are inevitably detected as an outside eavesdropper.

Besides, $P_1, P_2, \ldots, P_{a-1}, P_{a+1}, \ldots, P_n$ may hear of $r_2^a$ from $P_a$ in Step 5. According to Eq.(6), due to lack of $m_2^a$ and $k_a$, $P_1, P_2, \ldots, P_{a-1}, P_{a+1}, \ldots, P_n$ still cannot decode out $p_a$. In addition, $P_1, P_2, \ldots, P_{a-1}, P_{a+1}, \ldots, P_n$ may hear of $r_a$ from $TP_2$ in Step 6. However, it is useless for $P_1, P_2, \ldots, P_{a-1}, P_{a+1}, \ldots, P_n$ to get $p_a$. Finally, $P_1, P_2, \ldots, P_{a-1}, P_{a+1}, \ldots, P_n$ obtain the size relationship of different $p_i$ s from $TP_1$ in Step 7, where $i = 1, 2, \ldots, n$. Unfortunately, $P_1, P_2, \ldots, P_{a-1}, P_{a+1}, \ldots, P_n$ still has no access to $p_a$.

(3) The participant attack from semi-honest $TP_1$

In the proposed MQPC protocol, $TP_1$ is assumed to be semi-honest in the sense that she can misbehave at her own willing but is not allowed to conspire with anyone else including $TP_2$. When $P_i$ sends $r_2^i$ to $TP_2$ in Step 5, $TP_1$ may hear of $r_2^i$, where $i = 1, 2, \ldots, n$. However, $TP_1$ has no idea about $m_2^i$ and $k_i$ so that she still has no way to infer $p_i$ from $r_2^i$ according to Eq.(6). In addition, $TP_1$ knows $r_i$ from $TP_2$ in Step 6, but it is helpless for her to get $p_i$ in this Step. Finally, although $TP_1$ can calculate out $M_i$ in Step 7, she still has no way to obtain $p_i$.

(4) The participant attack from semi-honest $TP_2$

In the proposed MQPC protocol, $TP_2$ is supposed to be semi-honest, so she cannot conspire with $TP_1$. $TP_2$ gets $r_2^i$ from $P_i$ in Step 5, where $i = 1, 2, \ldots, n$. In order to infer $p_i$ from $r_2^i$, according to Eq.(6), $TP_2$ should know $m_2^i$ ahead. In order to achieve this goal, $TP_2$ may launch her attacks on $G_i$ during its transmission from $TP_1$ to $P_i$ in Step 3. However, $TP_2$ essentially acts as an outside attacker and is inevitably detected according to the analysis of Sect.4.1. As a result, $TP_2$ cannot get $m_2^i$ without being detected when launching her attacks. On the other hand, although $TP_2$ can know $m_1^i$, she still cannot get $m_2^i$ by virtue of Eq.(10), due to lack of $v_i$. In conclusion, $TP_2$ has no chance to obtain $p_i$ based on $r_2^i$ and $k_i$.

In addition, $TP_2$ may hear of the size relationship of different $p_i$ s from $TP_1$ in Step 7, but it is still helpless for her to acquire $p_i$.

## 5 Discussions

Ref.[36] defines the qudit efficiency as

$$\eta = \frac{z}{x+y}, \qquad (26)$$

where $x$, $y$ and $z$ are the number of consumed qudits, the length of consumed classical information and the length of compared private integers, respectively. In the following, we calculate the qudit efficiency of Eq.(26) for the proposed MQPC protocol. It needs to be declared that both the resources consumed for eavesdropping detection processes and the resources used for producing the pre-shared key $k_i$ are ignored, where $i = 1, 2, \ldots, n$.

In the proposed MQPC protocol, $P_i$ owns the private integer $p_i$, where $i = 1, 2, \ldots, n$, hence we have $z = 1$; $TP_1$ needs to prepare $n$ $d$-dimensional Bell states in Eq.(1), so we obtain $x = 2n$; besides, $P_i$ needs to send $r_2^i$ to $TP_2$, where $i = 1, 2, \ldots, n$, while $TP_2$ needs to send $R$ to $TP_1$, hence we acquire $y = n + n = 2n$. In short, the qudit efficiency of the proposed MQPC protocol is $\eta = \frac{1}{2n+2n} = \frac{1}{4n}$.

In addition, we compare the proposed MQPC protocol with previous MQPC protocols of size relationship in Refs.[31-35] in detail after ignoring the eavesdropping detection processes, and summarize the specific comparison outcomes in Table 1. According to Table 1, it can be concluded: ① As for quantum resources, the proposed MQPC protocol exceeds the protocols of Ref.[31] and Ref.[34], as $d$-dimensional Bell states are much easier to prepare than $d$-dimensional $n$-particle entangled states and $d$-dimensional GHZ states; ②As for the usage of unitary operation, the proposed MQPC protocol defeats the protocols of Refs.[32-34], as it doesn't need unitary operation; ③ the proposed MQPC protocol and the first protocol of Ref.[32] are the only two MQPC protocols of size relationship which require two TPs.

## 6 Conclusions

In this paper, in order to determine the size relationship of private integers from more than two users within one execution of protocol, we propose a novel MQPC protocol of size relationship with two semi-honest TPs by using $d$-dimensional Bell states, where two TPs are allowed to misbehave on their owns but cannot conspire with anyone else. The proposed MQPC protocol is proven to be secure against both the outside attacks and the participant attacks. The great merits of the proposed MQPC protocol lie in the following points:

(1) Two TPs only employ $d$-dimensional single-particle measurements rather than $d$-dimensional Bell state measurements;

(2) The proposed MQPC protocol needs neither quantum entanglement swapping nor unitary operations;

(3) The proposed MQPC protocol is feasible for the situation that $n$ users want to judge the size relationship of their private integers under the control of two supervisors;

(4) The proposed MQPC protocol can be used in the strange user environment, as there are not any communication and pre-shared key between each pair of users.

Recently, by introducing the concept of semiquantumness [37-39] into QPC, semiquantum private comparison (SQPC) [36,40-46] has gained considerable developments. However, few of the existing SQPC protocols can be used to determine the size relationship for private inputs from more than two classical users through being implemented one round. In the future, we will concentrate on studying this kind of SQPC.

Table 1  Comparison results of the proposed MQPC protocol with previous MQPC protocols of size relationship

| | Quantum resources | Number of parties | Number of TP | TP's quantum measurement | Users' quantum measurement | Comparison of size relationship | Usage of pre-shared key | Usage of unitary operation | Usage of quantum entanglement swapping |
|---|---|---|---|---|---|---|---|---|---|
| The protocol of Ref.[31] | $d$-dimensional $n$-particle entangled states | $n$ | 1 | No | $d$-dimensional single-particle measurements | Yes | Yes | No | No |
| The first protocol of Ref.[32] | $d$-dimensional single-particle states | $n$ | 2 | $d$-dimensional single-particle measurements | No | Yes | No | Yes | No |
| The | $d$- | $n$ | 1 | $d$- | No | Yes | Yes | Yes | No |
| second protocol of Ref.[32] | dimensional single-particle states | | | dimensional single-particle measurements | | | | | |
| The protocol of Ref.[33] | 2-dimensional single-particle states | $n$ | 1 | 2-dimensional single-particle measurements | | No | Yes | Yes | Yes | No |
| The protocol of Ref.[34] | $d$-dimensional GHZ states | $n$ | 1 | $d$-dimensional single-particle measurements | | No | Yes | No | Yes | No |
| The protocol of Ref.[35] | $d$-dimensional Bell states | $n$ | 1 | $d$-dimensional single-particle measurements | $d$-dimensional single-particle measurements | Yes | Yes | No | No |
| Our protocol | $d$-dimensional Bell states | $n$ | 2 | $d$-dimensional single-particle measurements | $d$-dimensional single-particle measurements | Yes | Yes | No | No |

## Acknowledgments

Funding by the National Natural Science Foundation of China (Grant No.62071430), the Fundamental Research Funds for the Provincial Universities of Zhejiang (Grant No.JRK21002) and the General Project of Zhejiang Provincial Education Department (Grant No.Y202250189) is gratefully acknowledged.

## Reference

[1] Yao, A.C.: Protocols for secure computations. In Proc. of the 23rd Annual IEEE Symposium on Foundations of Computer Science, pp.160-164 (1982)

[2] Yang, Y.G., Wen, Q.Y.: An efficient two-party quantum private comparison protocol with decoy photons and two-photon entanglement. J. Phys. A: Math. Theor. 42(5), 055305 (2009)

[3] Yang, Y.G., Gao, W.F., Wen, Q.Y.: Secure quantum private comparison. Phys. Scr. 80, 065002 (2009)

[4] Chen, X.B., Xu, G., Niu, X.X., Wen, Q.Y., Yang, Y.X.: An efficient protocol for the private comparison of equal information based on the triplet entangled state and single-particle measurement. Opt. Commun. 283, 1561-1565 (2010)

[5] Liu, W., Wang, Y.B., Cui, W.: Quantum private comparison protocol based on Bell entangled states. Commun. Theor. Phys. 57, 583-588 (2012)

[6] Tseng, H.Y., Lin, J., Hwang, T.: New quantum private comparison protocol using EPR pairs. Quantum Inf. Process. 11, 373-384 (2012)

[7] Yang, Y.G., Xia, J., Jia, X., Shi, L., Zhang, H.: New quantum private comparison protocol without entanglement. Int. J. Quantum Inf. 10, 1250065 (2012)

[8] Zi, W., Guo, F.Z., Luo, Y., Cao, S.H., Wen, Q.Y.: Quantum private comparison protocol with the random rotation. Int. J. Theor. Phys. 52, 3212-3219 (2013)

[9] Liu, B., Gao, F., Jia, H.Y., Huang, W., Zhang, W.W., Wen, Q.Y.: Efficient quantum private comparison employing single photons and collective detection. Quantum Inf. Process. 12, 887-897 (2013)

[10] Wang, C., Xu, G., Yang, Y.X.: Cryptanalysis and improvements for the quantum private comparison protocol using EPR pairs. Int. J. Quantum Inf. 11, 1350039 (2013)

[11] Yang, Y.G., Xia, J., Jia, X., Zhang, H.: Comment on quantum private comparison protocols with a semi-honest third party. Quantum Inf. Process. 12, 877-885 (2013)

[12] Zhang, W.W., Zhang, K.J.: Cryptanalysis and improvement of the quantum private comparison protocol with semi-honest third party. Quantum Inf. Process. 12, 1981-1990 (2013)

[13] Chen, X.B., Su, Y., Niu, X.X., Yang, Y.X.: Efficient and feasible quantum private comparison of equality against the collective amplitude damping noise. Quantum Inf. Process. 13, 101-112 (2014)

[14] Li, J., Zhou, H.F., Jia, L., Zhang, T.T.: An efficient protocol for the private comparison of equal information based on four-particle entangled W state and Bell entangled states swapping. Int. J. Theor. Phys. 53(7), 2167-2176 (2014)

[15] Liu, X.T., Zhang, B., Wang, J., Tang, C.J., Zhao, J.J.: Differential phase shift quantum private comparison. Quantum Inf. Process. 13, 71-84 (2014)

[16] Sun, Z.W., Long, D.Y.: Quantum private comparison protocol based on cluster states. Int. J. Theor. Phys. 52, 212-218 (2013)

[17] Ji, Z.X., Ye, T.Y.: Quantum private comparison of equal information based on highly entangled six-qubit genuine state. Commun. Theor. Phys. 65, 711-715 (2016)

[18] Ye, T.Y.: Quantum private comparison via cavity QED. Commun. Theor. Phys. 67(2),147-156 (2017)

[19] Ye, T.Y., Ji, Z.X.: Two-party quantum private comparison with five-qubit entangled states. Int. J. Theor. Phys. 56(5):1517-1529 (2017)

[20] Chang, Y.J., Tsai, C.W., Hwang, T.: Multi-user private comparison protocol using GHZ class states. Quantum Inf. Process. 12, 1077-1088 (2013)

[21] Wang, Q.L., Sun, H.X., Huang, W.: Multi-party quantum private comparison protocol with $n$-level entangled states. Quantum Inf. Process. 13, 2375-2389 (2014)

[22] Hung, S.M., Hwang, S.L., Hwang, T., Kao, S.H.: Multiparty quantum private comparison with almost dishonest third parties for strangers. Quantum Inf. Process. 16(2), 36 (2017)

[23] Ye, T.Y., Ji, Z.X.: Multi-user quantum private comparison with scattered preparation and one-way convergent transmission of quantum states. Sci. China Phys. Mech. Astron. 60(9), 090312 (2017)

[24] Ji, Z.X., Ye, T.Y.: Multi-party quantum private comparison based on the entanglement swapping of $d$-level Cat states and $d$-level Bell states. Quantum Inf. Process. 16(7), 177 (2017)

[25] Ye, C.Q., Ye, T.Y.: Circular multi-party quantum private comparison with $n$-level single-particle states. Int. J. Theor. Phys. 58, 1282-1294 (2019)

[26] Ye, T.Y., Hu, J.L.: Multi-party quantum private comparison based on entanglement swapping

of Bell entangled states within $d$-level quantum system. Int. J. Theor. Phys. 60(4), 1471-1480 (2021)

[27] Lin, S., Sun, Y., Liu, X.F., Yao, Z.Q.: Quantum private comparison protocol with $d$-dimensional Bell states. Quantum Inf. Process.12, 559-568 (2013)

[28] Yu, C.H., Guo, G.D., Lin, S.: Quantum private comparison with $d$-level single-particle states. Phys. Scr. 88, 065013 (2013)

[29] Guo, F.Z., Gao, F., Qin, S.J., Zhang, J., Wen, Q.Y.: Quantum private comparison protocol based on entanglement swapping of $d$-level Bell states. Quantum Inf. Process.12(8), 2793-2802(2013)

[30] Chen, F.L., Zhang, H., Chen, S.G., Cheng, W.T.: Novel two-party quantum private comparison via quantum walks on circle. Quantum Inf. Process. 20(5), 1-19 (2021)

[31] Luo, Q.B., Yang, G.W., She, K., Niu, W.N., Wang, Y.Q.: Multi-party quantum private comparison protocol based on $d$-dimensional entangled states. Quantum Inf. Process. 13, 2343-2352 (2014)

[32] Ye, C.Q., Ye, T.Y.: Multi-party quantum private comparison of size relation with $d$-level single-particle states. Quantum Inf. Process. 17(10), 252 (2018)

[33] Song, X.L., Wen, A.J., Gou, R.: Multiparty quantum private comparison of size relation based on single-particle states. IEEE Access 99, 1-7 (2019)

[34] Cao, H., Ma, W.P., Lü, L.D., He, Y.F., Liu, G.: Multi-party quantum comparison of size based on $d$-level GHZ states. Quantum Inf. Process. 18, 287 (2019)

[35] Wang, B., Gong, L.H., Liu,S.Q.: Multi-party quantum private size comparison protocol with $d$-dimensional Bell states. Front. Phys. 10, 981376 (2022)

[36] Geng, M.J., Xu, T.J., Chen, Y., Ye, T.Y.: Semiquantum private comparison of size relationship based $d$-level single-particle states. Sci. Sin. Phys. Mech. Astron. 52(9), 290311 (2022)

[37] Boyer, M., Kenigsberg, D., Mor, T.: Quantum key distribution with classical Bob. Phys. Rev. Lett. 99(14),140501 (2007)

[38] Ye, T.Y., Li, H.K., Hu, J.L.: Semi-quantum key distribution with single photons in both polarization and spatial-mode degrees of freedom. Int. J. Theor. Phys. 59, 2807-2815 (2020)

[39] Ye, T.Y., Geng, M.J., Xu, T.J., Chen, Y.: Efficient semiquantum key distribution based on single photons in both polarization and spatial-mode degrees of freedom. Quantum Inf. Process. 21, 123 (2022)

[40] Chou, W.H., Hwang, T., Gu, J.: Semi-quantum private comparison protocol under an almost-dishonest third party. https://arxiv.org/abs/1607.07961 (2016)

[41] Ye, T.Y., Ye. C.Q.: Measure-resend semi-quantum private comparison without entanglement. Int. J. Theor. Phys. 57(12),3819-3834 (2018)

[42] Thapliyal, K., Sharma, R.D., Pathak, A.: Orthogonal-state-based and semi-quantum protocols for quantum private comparison in noisy environment. Int. J. Quantum Inf. 16(5), 1850047 (2018)

[43] Lang, Y.F.: Semi-quantum private comparison using single photons. Int. J. Theor. Phys. 57, 3048-3055 (2018)

[44] Jiang, L.Z.: Semi-quantum private comparison based on Bell states. Quantum Inf. Process. 19, 180 (2020)

[45] Geng, M.J., Chen, Y., Xu, T.J., Ye, T.Y.: Single-state semiquantum private comparison based on Bell states. EPJ Quantum Technol. 9, 36 (2022)

[46] Ye, T.Y., Lian, J.Y.: A novel multi-party semiquantum private comparison protocol of size relationship with d-dimensional single-particle states. Physica A 611, 128424 (2023)